\pdfoutput=1
\documentclass{my-paper}

\usepackage{tikz}
\usetikzlibrary{arrows.meta}

\newcommand{\edgeglyph}[1][]{%
  \begin{tikzpicture}[baseline=-0.5ex]
    \draw[-Latex, semithick, #1] (0, 0) -- (1.2, 0);
  \end{tikzpicture}%
}

\newcommand{\boundaryglyph}{%
  \begin{tikzpicture}[baseline=-0.5ex]
    \draw[rounded corners=4pt, thick, gray] (0, -0.3) rectangle (1.2, 0.5);
  \end{tikzpicture}%
}

\title{The On-Chain and Off-Chain Mechanisms of
  \glstext*{dao}-to-\glstext*{dao} Voting\thanks{This is a preprint of
a published article~\citep{lloyd-et-al-24}.}}

\author{Thomas~Lloyd\thanks{\email{tomlloyd1992@gmail.com}}}

\author{Daire~\'O~Broin\thanks{\email{daire.obroin@setu.ie},
\orcid{0000-0002-2886-5546}}}

\author{Martin~Harrigan\thanks{\email{martin.harrigan@setu.ie},
\orcid{0000-0002-6069-7001}}}

\affil{\gls*{setu}, \acrlong*{roi}}

\date{}

\begin{document}

\maketitle

\begin{abstract}
  Voting is the primary mechanism through which \glsstar{}{dao}{s} reach
decisions.  Although transparent, the voting process can be opaque:
it can involve many interacting smart contracts.  The nexus of the
decision-making process can be relocated and the true voter
demographic obfuscated.  \glsstar{}{dao}{s} can also govern
other \glsstar{}{dao}{s}, a process known as \emph{metagovernance}.

We present a method for identifying \gls*{dao}-to-\gls*{dao}
metagovernance on the Ethereum blockchain.  We focus on the links
between \glsstar{}{dao}{s} and token contracts.  We use a
signature-matching algorithm to handle a variety of
\gls*{dao} frameworks and voting schemes.  Once we establish
token-to-\gls*{dao} relationships, we gather and process voting data
to produce a list of metagovernance relationships.  We apply this
algorithm to an initial set of sixteen \glsstar{}{dao}{s} and we
extend the dataset as more \glsstar{}{dao}{s} are identified.  We
produce a metagovernance network with \num{61} \glsstar{}{dao}{s} and
\num{72} metagovernance relationships.  We examine three case studies
that show metagovernance of various forms:  \emph{strategic},
\emph{decisive}, and \emph{nexus}, where a \gls*{dao} becomes a
governance hub for multiple other \glsstar{}{dao}{s}.

We demonstrate that metagovernance obscures voting context and
introduces entities driven by self-interest that can significantly
influence governance.  We highlight instances of metagovernance
between \glsstar{}{dao}{s} operating on the Ethereum blockchain where
current governance tools fail to reveal such dynamics.  Better tools
are needed to preserve the transparency-centric ethos of
\glsstar{}{dao}{s} and mitigate risks associated with
metagovernance.

\end{abstract}

\section{Introduction}\label{sec:introduction}

\glsstar{}{dao}{s} are optimistic in their construction: they rely on
voting to support collective decision-making.  Proponents claim that
\glsstar{}{dao}{s} offer community participation, transparency,
accountability, flexibility, and legitimacy.  Critics point to
bureaucratic inefficiencies, concentrations of power, the coalescence
of private interests, and voter
apathy~\citep{dupont-17,reijers-et-al-18,fritsch-et-al-22,sun-et-al-22,sun-et-al-22-1,kitzler-et-al-23}.
In this paper, we investigate an additional cause for concern:
\glsstar{}{dao}{s} routinely create and vote on proposals in other
\glsstar{}{dao}{s}\@.  Although this is widely known, its mechanisms
and significance remain largely unexplored.

In their simplest form, governance tokens represent voting rights in a
\gls*{dao} that are distributed to stakeholders.  Stakeholders vote on
proposals and, if a proposal passes, it is executed by the
\gls*{dao}\@.  \gls*{dao}-to-\gls*{dao} voting is a natural outcome: a
\gls*{dao} can acquire governance tokens of another \gls*{dao} and
exercise its voting rights.  This form of \emph{metagovernance} can
lead to complexities and tangled hierarchies.  While smart contract
composability is viewed favourably
within Decentralised Finance (DeFi), it may be problematic within
decentralised governance, where voting systems should be simple
and comprehensible to stakeholders.

We present a method to automatically detect instances of
metagovernance.  We gather an initial dataset comprising voting
contract addresses.  We refine prior methods by targeting
contracts directly linked to voting power contracts.  This approach
establishes connections between the governor and the voting power
source and it handles a wide variety of \gls*{dao} frameworks and voting
schemes.  Once we have identified the token-to-\gls*{dao}
relationship, we gather and process voting data.  This produces a list
of potential metagovernance relationships.  These instances are
labelled and visualised using a network graph.  Although we
demonstrate the method on a curated token list, it applies to any
governance token and can be reused with alternative indexing and
labelling sources.

The network graph shows observable governance influence in the
voting data; it does not reveal the internal decision-making process
of an external community.  The transactions of a
\num{2}-of-\num{3} multi-signature wallet show the outcome of an
external governance process, not the off-chain deliberation that
produced it, and the same applies to external communities voting
through Snapshot Spaces.  Further analysis of the voting data,
complemented by case studies, reveals three metagovernance dynamics.
We use the descriptive labels \emph{strategic}, \emph{decisive}, and
\emph{nexus} for these observed behaviours; they are illustrative
examples rather than a complete taxonomy.

This paper is organised as follows.  In \cref{sec:related-work} we
review related work, specifically smart contract identification
schemes and the governance risks associated with \glsstar{}{dao}{s}\@.
We define metagovernance and its variations in
\cref{sec:metagovernance}, describe the primary components of a
\gls*{dao} governance framework, and introduce a visual grammar for
governance diagrams.  We describe an algorithmic approach
to detect metagovernance using on-chain and off-chain data in
\cref{sec:method}.  \Cref{sec:results} summarises our results and
presents a network visualisation of metagovernance between
\glsstar{}{dao}{s}\@.  We consider three case studies in
\cref{sec:case-studies}.  Finally, \cref{sec:conclusions} concludes the paper
and proposes future work.

\section{Related Work}\label{sec:related-work}

We divide related work into two categories: smart contract
identification schemes and the risks and trade-offs with respect to
blockchain-based governance.

Although smart contracts can contain arbitrary code, they can be
classified into distinct groups.  \citet{bartoletti-pompianu-17}
present a taxonomy of smart contracts, manually inspecting their code
and classifying them into distinct groups.  They further categorised
design patterns shared across smart contracts.

\citet{frowis-et-al-19} propose a dual-method approach to
automatically identify token smart contracts.  First, they analyse
bytecode execution paths (traces) and the location of on-chain storage
data manipulation.  Second, they match function signatures to the
ERC-20 standard.  They use signature matching to good effect while
cautioning that it relies on standard implementations of token smart
contracts.

\citet{he-et-al-20} highlight the homogeneity of smart contracts on
the Ethereum blockchain by generating a fuzzy hash from pre-processed
smart contract bytecode and checking for similarities using pair-wise
comparison.

\glsstar{}{dao}{s} have governance risks.  \citet{dhillon-et-al-17}
and \citet{dupont-17} review \textquote{The DAO} and the smart
contract code that left it exploitable.  They discuss the
repercussions and the fallout from the exploit.
\citet{feichtinger-et-al-24} identify risk factors across several
\glsstar{}{dao}{s} and present strategies to mitigate these risks,
such as the development of new governance tools.

\citet{lloyd-et-al-23} consider the outcomes of the veToken governance
model: it leads to metagovernance protocols holding a significant
amount of voting power and the creation of bribe markets that have
impactful outcomes on proposals.

\citet{rossello-24} demonstrates that economic returns in token
valuation follow strategic voter decisions.  He observes the different
outcomes in two types of voters: \emph{large voters}, who can make
unilateral decisions with their voting power, produce negative returns
while \emph{blockholders}, who hold significant voting power without
being able to make unilateral decisions, produce positive returns.

\citet{austgen-et-al-23} propose a method to analyse centralisation
in \glsstar{}{dao}{s} that goes beyond conventional address balance
analysis:  shared utility incentives.  They demonstrate the risk of
anonymous bribes through a \textquote[][.]{Dark \gls*{dao}} It shields
the recipients from scrutiny by using a trusted execution environment
to hide their activities.

\citet{kitzler-et-al-23} combine data from Ethereum, Snapshot and
third-party APIs to investigate the centralised position of
contributors, i.e., project owners, administrators, and developers, in
\gls*{dao} governance structures.  They show that, for most
\glsstar{}{dao}{s}, the relative voting power of contributors is low
but in some \glsstar{}{dao}{s}, especially ones with a high total
value locked, it is high.  Additionally, contributors tend to exercise their voting power
and they vote similarly to other contributors in the same
\glsstar{}{dao}{s}.  The authors also found instances of large
token transfers in temporal proximity to the closing of polls,
although this is not fully explained.

The contract-identification work above is concerned with classifying
contracts in isolation, while the governance-risk work focuses on
voter behaviour or token concentration within a single \gls*{dao}.
Neither line of work addresses the relationships \emph{between}
\glsstar{}{dao}{s} that arise when one \gls*{dao} holds the
governance token of another.  Our contribution combines signature
matching with on-chain trace data to identify these inter-\gls*{dao}
relationships and to characterise the resulting metagovernance
structures.

\section{Metagovernance}\label{sec:metagovernance}

The term \emph{metagovernance} was coined in 1997~\citep{jessop-97} in
the context of governments: they set basic rules, ensure different
systems work together, resolve disputes, aim to balance power among
groups, influence how organisations see themselves and their
strategies, and take responsibility when these governance efforts fall
short.  Metagovernance is an
\textquote[\citep{van-bortel-mullins-09}][.]{attempt by politicians,
  administrators or other governance networks to construct, structure
and influence the game-like interaction within governance networks}
It is the governance of governance.

\begin{figure}[htbp]
  \centerline{\myincludegraphics[width=\linewidth]{png/figures/metagovernance}}
  \caption{Metagovernance occurs between two \glsstar{}{dao}{s},
    \gls*{dao}~A and \gls*{dao}~B, when the result of a proposal in
    \gls*{dao}~A either creates or votes on a proposal in
    \gls*{dao}~B.  The solid lines show the progression between the
    stages in a typical proposal process within \gls*{dao}~A and
    \gls*{dao}~B; the dotted lines show the output of the process in
    \gls*{dao}~A becoming an input to the process in
  \gls*{dao}~B.}\label{fig:metagovernance}
\end{figure}

Metagovernance between \glsstar{}{dao}{s} is an attempt by one
\gls*{dao} to govern another.  \Cref{fig:metagovernance} illustrates
one form of metagovernance between two \glsstar{}{dao}{s},
\gls*{dao}~A and \gls*{dao}~B.  A proposal in \gls*{dao}~A progresses
through the typical stages:  creation, voting, tallying and execution.
The progression is represented by solid lines.  The outcome of the
proposal either creates or votes on a proposal in \gls*{dao}~B.  The
output of the former proposal is an input to the latter; this is
represented by dotted lines.  In the following subsections, we
highlight variations on this form of metagovernance between
\glsstar{}{dao}{s}.

\subsection{On-Chain/Off-Chain}

Typically, the stages in \cref{fig:metagovernance} (proposal creation,
voting, tallying and execution) occur on-chain.  However,
\glsstar{}{dao}{s} can perform some of the stages off-chain if, for
example, the stakes are low, on-chain transaction fees are high, or if
the proposal is a \textquote{temperature check} for a future proposal.
\citet{snapshot-xx} is a popular off-chain voting application that
supports this.  To identify intra-\gls*{dao} governance (the
solid lines) and inter-\gls*{dao} governance or metagovernance (the
dotted lines), our method in \cref{sec:method} combines on-chain
and off-chain data.

\subsection{Voting Power}

In \cref{fig:metagovernance}, how might \gls*{dao}~A obtain voting
power in \gls*{dao}~B?  Typically, \gls*{dao}~A has \emph{tokenised
voting power}:  it holds the governance token of \gls*{dao}~B and
therefore receives a proportional amount of voting rights.  However,
there are alternatives such as \emph{token-locked voting power} where
\gls*{dao}~A must lock the governance token of \gls*{dao}~B for a time
period in exchange for voting rights~\citep{lloyd-et-al-23}.  The
intention is to align the, potentially short-term, interest of the
token holders with the long-term goals of the \gls*{dao} by requiring
the token holders to commit to a lockup period.  \emph{Council voting
power} is yet another alternative where \gls*{dao}~A is elected as a
member of a governing council for \gls*{dao}~B.

\gls*{dao}~A can obtain \gls*{dao}~B's governance token in many
ways: on the open market, through token swaps, as a reward for
contributions, collaborations, or liquidity provision.  Our method
in \cref{sec:method} focuses on voting power rather than governance
tokens to sidestep much of this complexity.

\subsection{Delegation}

Delegation enables an entity with voting rights to delegate those
rights to another entity.  It allows passive entities to trust
\textquote{\gls*{dao} politicians} to act in their best interest.
Many of the \glsstar{}{dao}{s} we study in this paper delegate their
voting rights.  For example, \gls*{dao}~A's proposal in
\cref{fig:metagovernance} could be to delegate its voting rights in
\gls*{dao}~B to a \gls*{dao} politician.  Then, the \gls*{dao}
politician either creates or votes on a proposal in \gls*{dao}~B.  We
examine this configuration in \cref{sec:results}.  There is also the
possibility that voting rights are delegated directly \emph{to} a
\gls*{dao}, i.e., the \gls*{dao} becomes the politician.  In practice,
this rarely occurs.

\subsection{Governance Frameworks}

\gls*{dao} governance frameworks vary significantly as each \gls*{dao}
customises its structure to meet its specific needs and objectives,
including when a \gls*{dao} decides to influence another
\glsstar{}{dao}{'s} governance.  This process adds complexity to the
governance ecosystem, as the \emph{metagovernor} must follow the rules
of the \emph{metagovernee} it aims to influence while also setting its
own rules to serve its interests.  Metagovernance introduces both
opportunities and risks, making it important to detect and understand
its implications.

\subsection{Visual Grammar for Governance
Diagrams}\label{sec:visual-grammar}

To describe governance structures precisely, we define a visual
grammar (\cref{tab:visual-grammar}) comprising five vertex types,
five edge types, and one annotation convention.  We use this grammar
to illustrate the governance scenarios in our analysis, particularly
the case studies in \cref{sec:case-studies}.  Where multiple
instances of the same type appear in a diagram, they are
distinguished by labels (e.g., Token~A and Token~B, or \gls*{dao}~A
and \gls*{dao}~B).

\begin{table}[htbp]
  \centering
  \renewcommand{\arraystretch}{1.2}
  \caption{The visual grammar for governance diagrams comprises
    five vertex types, five edge types, and one annotation
    convention.}\label{tab:visual-grammar}
  \begin{tabular}{>{\centering\arraybackslash}m{1.6cm}>{\raggedright\arraybackslash}m{3.6cm}m{8.8cm}}
    \toprule
    \multicolumn{3}{l}{\emph{Vertices}} \\
    \midrule
    \myincludegraphics[width=0.9cm]{svg/people/person} &
    Actor &
    A person or entity that participates in governance, possibly
    controlling one or more blockchain addresses. \\
    \myincludegraphics[width=0.9cm]{svg/objects/wallet} &
    \glstext*{eoa} &
    An externally-owned account: a blockchain address controlled by
    a single private key. \\
    \myincludegraphics[width=0.9cm]{svg/objects/token} &
    Token &
    A governance token.  Holding the token confers voting power in
    the associated \gls*{dao}\@. \\
    \myincludegraphics[width=1.2cm]{svg/objects/multisig} &
    Multi-signature wallet &
    A smart contract address requiring $m$-of-$n$ signers to approve
    a transaction. \\
    \myincludegraphics[width=1.2cm]{svg/objects/dao} &
    \glstext*{dao} &
    A governance contract: the hub where proposals are submitted and
    votes are cast. \\
    \midrule
    \multicolumn{3}{l}{\emph{Edges}} \\
    \midrule
    \edgeglyph[double, double distance=1.5pt] &
    Controls &
    A double-lined arrow indicating that the source actor holds the
    private key for the target address.  For a multi-signature
    wallet, multiple control edges converge on a single vertex. \\
    \edgeglyph &
    Holds &
    A solid arrow indicating that the source address holds the
    target token, conferring voting power. \\
    \edgeglyph[dashed, T-Q-M4] &
    Votes yes &
    A dashed green arrow indicating a vote cast in support of a
    proposal. \\
    \edgeglyph[dashed, red!60!black] &
    Votes no &
    A dashed red arrow indicating a vote cast in opposition to a
    proposal. \\
    \edgeglyph[dash dot] &
    Tokenised by &
    A dash-dot arrow indicating a tokenisation relationship: the
    source token is deposited into a contract that mints the target
    token. \\
    \midrule
    \multicolumn{3}{l}{\emph{Conventions}} \\
    \midrule
    \boundaryglyph &
    Community boundary &
    A grey rounded rectangle grouping actors that share governance
    alignment or organisational membership. \\
    \bottomrule
  \end{tabular}
\end{table}

The five vertex types capture the distinct roles that entities play
in a governance system.  An \emph{actor} is a person or entity that
participates in governance.  We use \textquote{actor} rather than
\textquote{owner} because not all governance participants own
tokens: some receive delegated voting power, and others act through
multi-signature wallets they share with other actors.  Actors
interact with a blockchain through addresses.  An
\glsstar{}{eoa}{} is a standard blockchain address controlled by a
single private key.  A \emph{multi-signature wallet} is a smart
contract address that requires $m$-of-$n$ designated signers to
approve a transaction before it can execute.  The distinction
between these two address types matters for our method
(\cref{sec:method}): many governance voters are contract accounts
rather than \glsstar{}{eoa}{s}, and detecting multi-signature
wallets is a key step in identifying external communities.  A
\emph{token} is a governance token whose balance determines voting
power in a \gls*{dao}\@.  A \emph{\gls*{dao}} is the governance
system itself: the contract (or off-chain Space) where proposals are
submitted and votes are cast.

The five edge types represent the relationships between vertices.  A
\emph{controls} edge connects an actor to an address, indicating
that the actor holds the private key for that \gls*{eoa} or is one
of the designated signers of that multi-signature wallet.  A
\emph{holds} edge connects an address to a token, indicating that
the address has a token balance and therefore voting power.  Two
vote edges capture governance actions: a \emph{votes yes} edge
indicates a vote cast in support of a proposal and a \emph{votes no}
edge indicates opposition.  In both cases, the voting power
exercised by the address derives from the tokens it holds.  A
\emph{tokenised by} edge connects one token to another, indicating
that holders of the source token deposit it into a smart contract
that mints the target token; the edge is directed from the
underlying token to the derivative.

One annotation convention completes the grammar.  A \emph{community
boundary} is a rounded rectangle drawn around a group of actors that
share governance alignment or organisational membership.  A
community boundary may also serve as the source or target of an
edge, indicating that the action is performed collectively by the
community rather than by any single actor.

The grammar constrains which edges may connect which vertex types.
Actors connect only to addresses via controls edges; addresses
connect to tokens via holds edges and to \glsstar{}{dao}{s} via vote
edges; tokenised by edges connect one token to another.  Tokens
never vote; only addresses do.  Every governance action therefore
passes through an address, reflecting the structure of the
underlying blockchain.

\Cref{fig:visual-grammar-example} illustrates the grammar with an
instance of tokenised metagovernance.  Token~B is tokenised by
Token~A, creating a new governance layer (\gls*{dao}~A) on top of
the underlying \gls*{dao}~B\@.  Two communities within \gls*{dao}~A
vote on a proposal, and a multi-signature wallet controlled by
\gls*{dao}~A enacts the outcome in \gls*{dao}~B\@.  The locus of
governance shifts to the higher tier: the vote recorded in
\gls*{dao}~B is a single yes from the multi-signature wallet, while
the deliberation that produced it took place in \gls*{dao}~A\@.  In
the following section we describe our method for identifying
\gls*{dao} metagovernance.

\begin{figure}[htbp]
  \centerline{\myincludegraphics[width=0.7\linewidth]{tikz/dao-governance/dao-governance-metagovernance}}
  \caption{Tokenised metagovernance expressed in the visual grammar.
    Token~B is tokenised by Token~A, creating a new governance layer
    (\gls*{dao}~A) on top of the underlying \gls*{dao}~B\@.
    Communities within \gls*{dao}~A vote on proposals, and a
    multi-signature wallet controlled by \gls*{dao}~A enacts the
    outcome in \gls*{dao}~B\@.}\label{fig:visual-grammar-example}
\end{figure}

\section{Method}\label{sec:method}

To investigate \gls*{dao}-to-\gls*{dao} voting mechanisms, we
collected on-chain and off-chain data.  We extracted data from
Ethereum mainnet using an Ethereum archive node, including event logs,
trace calls and historical contract calldata, from block height
\numrange{0}{18299999} (October 2023) inclusive.  We collected
off-chain data from a variety of sources, including voting data from
the Snapshot GraphQL \glstext*{api}, labelled event signatures from
\citet{etherface-xx}, smart contract Application Binary Interfaces
(ABIs) and labelled contract data from Etherscan, and a \gls*{dao}
token list from \citet{coinmarketcap-xx}.

The \gls*{dao} token list was a seed dataset of twenty tokens with the
highest market capitalisation and labelled with \textquote{Ethereum}
and \textquote{\glstext*{dao}} by CoinMarketCap.  We refined our list to sixteen
tokens primarily governed via the Ethereum blockchain.  We then
cross-referenced the entities between the different sources, matching
each token on the \gls*{dao} token list with its corresponding
governance smart contract on Ethereum mainnet and its corresponding
Space on Snapshot.  We detail this process in the following
subsections.

\subsection{Governance Smart Contracts}

Our first task was to match the set of \gls*{dao} governance tokens
with their governance smart contracts.  This is typically a manual
process that requires reading a protocol's documentation and GitHub
repository and inspecting its on-chain deployments.  We created an
automated process that combines on-chain and off-chain data to produce
similar results.  \textsc{IdentifyGovernanceSmartContract} identifies
the governance smart contract associated with a contract account
address (\cref{alg:gov-smart-contract-heuristic}).  The input
contract account address ($A$) can point to a governance token, or to
any source of voting power.  The output is the governance contract
account address that wields voting power over $A$.

\begin{algorithm}[H]
  \caption{The Governance Smart Contract Identification Heuristic
    (\textsc{IdentifyGovernanceSmartContract}) takes a contract
    account address ($A$) as input and produces a governance contract
  account address as output.}
  \label{alg:gov-smart-contract-heuristic}
  \begin{algorithmic}[1]
    \Function {IdentifyGovernanceSmartContract}{$A$}
    \State $B \gets
    \Call{ContractInvocators}{A}$\label{lin:contract-invocators}
    \State $C \gets \{\}$\label{lin:start-filtered-contracts}
    \ForAll{$b \in B$}
    \If{$\Call{HasEtherscanABI?}{b}$}\label{lin:etherscan-api}
    \State $ES \gets
    \Call{HashedEventSignatures}{b}$\label{lin:hashed-events}
    \If{$\Call{IsGovernanceContract?}{ES}$}
    \State $C \gets C \cup \{b\}$
    \EndIf
    \EndIf
    \EndFor\label{lin:end-filtered-contracts}
    \State \textbf{return}
    $\Call{MostFrequentlyInvocated}{C,
    A}$\label{lin:most-freq-invocated}
    \EndFunction
  \end{algorithmic}
\end{algorithm}

The algorithm first identifies the set of contract accounts that call
functions on the input contract address
(Line~\ref{lin:contract-invocators}).  These are obtained from the
historical trace data, specifically the \texttt{to} field of contract
calls directed at the input contract.  Governance systems based on
token voting typically separate the governance contract, which
manages proposals and votes, from the voting power contract, which
records token balances.  When a vote is cast, the governance contract
calls the voting power contract to determine the caller's voting
weight.  In one-token-one-vote frameworks, this is most commonly done
via the \texttt{balanceOf} function, which returns the number of
tokens held by a given address.  By starting from the voting power
contract and following the trace data backwards, we surface the
governance contracts that consult it.

Next, we filter the set of contract addresses to those that appear to
be governance contracts
(Lines~\ref{lin:start-filtered-contracts}--\ref{lin:end-filtered-contracts}).
For each contract on our list, we retrieved its ABI from the
Etherscan \glstext*{api} (Line~\ref{lin:etherscan-api}).  The ABI
gives us complete access to the contract's functions and events, and
we hashed the events linked to each contract
(Line~\ref{lin:hashed-events}).  We then matched these hashes against
a local database of known governance event signatures, built from
Etherface and \citet{4byte-xx}, restricted to signatures whose names
contain governance-related terms such as \textquote{vote} or
\textquote{proposal}.  This produces a set of candidate governance
contracts ($C$).

Finally, we select the candidate that most frequently invoked
functions on the input contract address as the governance contract
(Line~\ref{lin:most-freq-invocated}).

\begin{algorithm}[H]
  \caption{The Multi-Signature Contract Account Identification
    Heuristic (\textsc{IdentifyMultiSigVoters}) takes a governance
    contract account ($C$) as input and produces a set of
  multi-signature contract accounts as output ($W$).}
  \label{alg:multisig-voters-heuristic}
  \begin{algorithmic}[1]
    \Function {IdentifyMultiSigVoters}{$C$}
    \State $ES \gets \Call{HashedVotingSignature}{C}$\label{lin:hashed-vote}
    \State $V \gets \Call{ContractVoters}{C, ES}$\label{lin:voters}
    \State $W \gets \{\}$
    \ForAll{$v \in V$}
    \If{$\Call{HasMultiSig?}{v}$}
    \State $W \gets W \cup \Call{GetMultiSig}{v}$
    \EndIf
    \EndFor
    \State \textbf{return} $W$
    \EndFunction
  \end{algorithmic}
\end{algorithm}

Our second task was to navigate from a governance contract account to
the multi-signature contract accounts that voted in its governance
process.  We created an automated process to achieve this.
\textsc{IdentifyMultiSigVoters} identifies the contract account voters
for a single governance contract account address
(\cref{alg:multisig-voters-heuristic}).  The input governance contract
account address ($C$) is the output of the previous algorithm.  The
output is a set of contract account voters ($W$) that may be engaged
in metagovernance.

We first collect all event logs emitted by the contract and order
them by frequency, with the most frequent first.  We assume that the
most frequently occurring event log on a governance contract is
related to voting (Line~\ref{lin:hashed-vote}), since governance
systems generally produce more vote actions than proposal
submissions.  We then focus on events whose topic hash matches our
criteria for voting events.  We look for log entries labelled
\textquote{voter} that contain an Ethereum address; if no such
entries exist, we examine the \texttt{data} section of each log for
Ethereum addresses.  Once we have identified all addresses associated
with voting, we filter to those that are contract accounts
(Line~\ref{lin:voters}), leaving a list of contract accounts that
participated in the governance process.

\subsection{Delegation Data}

We identify delegation for on-chain contract accounts using a
procedure similar to that of \textsc{IdentifyGovernanceSmartContract}
(Lines~\ref{lin:etherscan-api}--\ref{lin:hashed-events}), and resolve
the delegation with a function called \texttt{GetDelegateSignature}.
The voting power address serves as input, and the function performs
signature matching against the keyword \textquote{delegate}.  Unlike
\textsc{IdentifyGovernanceSmartContract}, which returns a single
contract, \texttt{GetDelegateSignature} returns all
delegation-related signatures.  We then parse the delegate logs to
extract contract account delegations.

\subsection{Snapshot Spaces}\label{sec:snapshot-spaces}

To detect Snapshot Spaces from a source of voting power, we indexed
all Spaces and Strategies using the Snapshot GraphQL
\glstext*{api}\@.  This allowed us to look up a source of voting
power, such as a governance token address, and search for any
Snapshot Space that referenced it as part of its Strategy.  We
restricted the list of Snapshot Spaces to those with at least
\num{50} followers, since anyone can create a Space that references
any governance token.  We then retrieved all historical voting data
from the Snapshot \glstext*{api}, focusing exclusively on voting
addresses that were contract accounts.

\subsection{Account Labelling}

We labelled the list of contract account voter addresses using
Etherscan's public name tags.  Public name tags attach a
human-readable identifier to a smart contract address, including
identifiers for multi-signature accounts, treasuries, and even known
exploiters.

\section{Results}\label{sec:results}

In our analysis of \num{16} \glsstar{}{dao}{s}, we found \num{12} that
use an on-chain governance framework.  Of these, we identified \num{9}
correctly and one incorrectly.  Our method was unable to recognise
\citet{sky-xx} because the structure of its event logs differs from
the conventional pattern, producing a mismatch during the event log
signature step.  Although we identified Yearn's governance structure
from our Snapshot data, we erroneously tagged the Yearn locker as the
governance contract because the locker received a disproportionately
large number of trace calls.  We also encountered difficulty parsing
voting data for UMA, which uses an unconventional vote-revelation
scheme that is not easily detected through log counts.  Half of the
\num{16} \glsstar{}{dao}{s} draw voting power from sources other than
the token contract (for example, Curve's \unit{\vecrv}), which
required manual verification of the voting power source.  In total, we
identified \num{663} contract accounts participating in governance
operations.  Of these, we labelled \num{42} using Etherscan's tag
system, and we labelled five additional \glsstar{}{dao}{s} via
Snapshot Spaces.

We expanded the dataset by applying the same algorithm to the tagged
\glsstar{}{dao}{s} with governance tokens, iterating to a depth of
three and excluding any \glsstar{}{dao}{s} not engaged in governance
on Ethereum mainnet.  This process correctly identified two further
on-chain addresses and incorrectly identified four.  For Convex, for
example, our system flagged the Pirex \unit{\cvx} contract as a
governance contract: Pirex \unit{\cvx} is a token wrapper for
\unit{\vlcvx} that interacts directly with the \texttt{cvxLocker} and
emits vote logs relevant to Votium, which our heuristic misinterpreted
as a governance role.  We also found \num{11} additional Snapshot
addresses, of which only two contained no contract account voting.

We identified the corresponding Snapshot Space for every \gls*{dao} in
the original \num{16}.  This was expected: the voting power contract
for each had been manually derived from its documentation, and the
follower-count filter (\cref{sec:snapshot-spaces}) discarded Spaces
that referenced a token without a real community behind them.  Because
our dataset focused on \glsstar{}{dao}{s} with significant market
capitalisation, the algorithm might face difficulty with lesser-known
\glsstar{}{dao}{s}.

Our success rate in detecting on-chain governance was
\qty{61}{\percent}: of the \num{18} \glsstar{}{dao}{s} we attempted to
process end-to-end, we automated governance contract detection, log
parsing, and contract account voter retrieval for \num{11}.  The
remaining \qty{39}{\percent} (\num{7} \glsstar{}{dao}{s}) divides into
two groups.  Two were on-chain misidentifications caused by
inaccuracies in our algorithm's trace ordering and signature matching,
where wrapper contracts such as the Yearn locker and the Pirex
\unit{\cvx} wrapper were flagged in error.  The other
\qty{28}{\percent} (\num{5} \glsstar{}{dao}{s}) failed at the Snapshot
stage: although our pipeline returned a Space for each, no contract
account voting could be extracted from those Spaces, leaving no
metagovernance edges to add.

\subsection{Metagovernance Network}

\begin{figure}[htbp]
  \centerline{\myincludegraphics[width=\linewidth]{png/figures/metagovernance-network}}
  \caption{The metagovernance relationship between \glsstar{}{dao}{s}
    can be visualised as a network: each vertex corresponds to a
    \gls*{dao} and each directed edge connects a source to a target
    where the \gls*{dao} corresponding to the source metagoverns, or
    influences the governance of, the \gls*{dao} corresponding to the
  target.}\label{fig:metagovernance-network}
\end{figure}

\Cref{fig:metagovernance-network} visualises metagovernance as a
network.  Each vertex represents a \gls*{dao}, and each directed edge
from a source to a target represents the source \gls*{dao} exerting
influence on the target \gls*{dao}\@.  We identified \num{61} labelled
smart contracts, each represented as a vertex.  Convex Finance has the
highest in-degree, receiving participation in its governance from
\num{12} other metagovernors.  Curve Finance and Balancer follow, each
with \num{10}.

The three \glsstar{}{dao}{s} most heavily involved in metagovernance
all govern token emissions, which suggests that economic incentives
tied to gauge proposal outcomes attract metagovernors.  This
concentration may also reflect a bias in our method: contract account
voting on gauge contracts requires whitelisting, which may produce
more thorough labelling of metagovernors on Etherscan.  Convex Finance
complicates this picture: it does not require whitelisting for
participation in its own governance, yet has accumulated a central
role in Curve Finance's governance through an off-chain framework.

The vertex with the highest out-degree is GFX Labs, a blockchain
research and development organisation that participates in \gls*{dao}
governance, receives substantial delegations, and frequently creates
proposals.  Three \glsstar{}{dao}{s} (Synthetix, Index Coop, and
Redacted) share second position in out-degree.  Synthetix operates a
governance council, and Redacted, a liquidity aggregator, uses its
liquidity to acquire governance positions.  Index Coop is the subject
of a case study in \cref{sec:case-study-index-coop}.

\subsection{Delegation}

\begin{figure*}[htbp]
  \centerline{\myincludegraphics[width=\linewidth]{tikz/dao-governance/metagovernance-delegation}}
  \caption{\glsstar{}{dao}{s} delegate voting rights to third parties
    who vote on their behalf.  Although our dataset included
    \num{1299} instances of contract accounts delegating their voting
    rights, this only added \num{12} additional edges to the
    metagovernance network in \cref{fig:metagovernance-network}.  At
    each fork, the two branches have widths proportional to the square
    root of their delegate counts, normalised so that they sum to the
    width of the parent.  This scaling, applied recursively from root
    to leaves, preserves the ordering and approximate relative
    magnitude at each split while keeping the smallest leaves (e.g.,
    \textquote{Known Edges} with \num{4} delegates) legible alongside
  the initial flow of \num{1299}.}\label{fig:delegation-sankey}
\end{figure*}

\Cref{fig:delegation-sankey} visualises the filtering of on-chain
delegations into new metagovernance edges, read from left to right.
Our analysis identified six on-chain \glsstar{}{dao}{s} that use
delegation, across two distinct frameworks.  The first requires token
holders to delegate to themselves (\emph{self-delegation}) before they
receive governance rights.  The second distributes voting rights
without the need for self-delegation.  Of the \num{1299} contract
account delegations we observed, \num{475} were self-delegations and
\num{824} were not.

The two flows merge at a labelling step.  Using Etherscan's tag
system, we labelled \num{78} of the \num{1299} delegating addresses
and discarded the remaining \num{1221} as unlabelled.  Grouping the
labelled addresses by their parent \gls*{dao} yielded \num{38}
distinct \gls*{dao}-to-\gls*{dao} delegations; the \num{40} addresses
that did not aggregate into a \gls*{dao} are excluded.  Inspection of
the voting logs showed that only \num{16} of the \num{38} delegates
had cast votes, with the other \num{22} dropped.  Of the \num{16}
active delegates, four were already edges in our metagovernance
network and the remaining \num{12} were added as new edges
(\cref{fig:metagovernance-network}).

\section{Case Studies}\label{sec:case-studies}

We analysed three case studies drawn from our metagovernance network,
each chosen to illustrate a different metagovernance dynamic.  As in
\cref{sec:introduction}, the labels \emph{strategic}, \emph{decisive},
and \emph{nexus} describe the observed behaviours rather than a
complete taxonomy.  The first illustrates strategic metagovernance: a
single high-stakes vote in which the metagovernor's voting power
exceeded \qty{15}{\percent}.  The second illustrates decisive
metagovernance: a vote that directly determined a proposal's outcome.
The third illustrates nexus metagovernance: a \gls*{dao} that acts as
a governance hub for multiple other \glsstar{}{dao}{s}, evidenced by a
high out-degree in the metagovernance network and frequent references
to the governed \glsstar{}{dao}{s} in its own proposals.

\subsection{Liquity and Aave}\label{sec:case-study-liquity}

Liquity is a decentralised lending protocol and the issuer of the LUSD
stablecoin~\citep{liquity-xx}.  Aave is a decentralised lending
protocol for both stable and variable on-chain assets~\citep{aave-xx}.
Liquity cast a single vote with \qty{100000}{\aave} tokens, or
\qty{17}{\percent} of the total voting power, on Aave
\glsstar{}{dao}{'s} Proposal~\num{95}.  The proposal was first raised
on 3rd June 2022 in the Aave governance forum with the title
\textquote[][.]{ARC: Add support for LUSD (Liquity)} ARC is a
permissioned version of the Aave protocol designed for
anti-money-laundering and know-your-customer compliance, so tokens
must be whitelisted to appear on the platform; new assets are listed
only after a successful governance proposal.

The proposal transitioned to a Snapshot governance vote on 5th July
and concluded on 10th July with a \qty{94}{\percent} approval.
Liquity's direct involvement with the on-chain proposal began with a
delegation of \unit{\aave} tokens to its \textquote{Liquity: Bounties}
contract account.  On 16th August, an \gls*{eoa} delegated
\qty{100000}{\aave} tokens to the Liquity contract; nine days later,
the contract submitted Proposal~\num{95} to the governance contract
and cast a \textquote{yes} vote later the same
day.\footnote{\url{https://archive.today/sOL8o}} The proposal passed.

The proposal had direct benefits for Liquity, as it expanded the
utility of LUSD\@.  However, the rapid accumulation of voting power to
influence the token's listing raises concerns characteristic of
strategic metagovernance.

\subsection{Blockchain at Berkeley and Compound
Finance}\label{sec:case-study-blockchain-at-berkeley}

Blockchain at Berkeley was founded in 2016.  They describe themselves
as \textquote[\citep{blockchain-at-berkeley-xx}][.]{driving innovation
  in the blockchain industry by building an ecosystem that empowers
  students to make an impact through practical education, consulting
for enterprise companies, and conducting open source research}
Within our metagovernance network (\cref{fig:metagovernance-network}),
Blockchain at Berkeley participates in the governance of Aave,
Compound, and Uniswap.  They have voted \num{70} times across these
three \glsstar{}{dao}{s}, accounting for almost \qty{6}{\percent} of
the total votes cast per proposal on average.

Compound Finance~\citep{compound-xx} is a decentralised lending
protocol.  On 15th April 2022, Proposal~\num{100},\footnote{See
\url{https://compound.finance/governance/proposals/100}.} titled
\textquote{COMP Rewards Adjustments: Kickstart Rewards: Step Two}, was
created within Compound's \gls*{dao}\@.  The proposal sought to end
the COMP rewards programme and replace it with a new programme focused
on bootstrapping new markets.  Voting ran from 18th to 21st April.
The proposal was rejected, with \num{492678} votes in favour and
\num{499849} against.  Blockchain at Berkeley cast the deciding vote
of \num{50007} against.\footnote{See
\url{https://archive.today/wUtjJ}.}

By focusing on votes cast by contract addresses that were decisive in
proposal outcomes, our analysis reveals the influence that a single
external community can exert within a \gls*{dao}\@.  Blockchain at
Berkeley operates through a multi-signature wallet\footnote{See
\url{https://archive.today/Tfzqh}.} that requires authorisation from
two of its eleven owners; the off-chain process by which those signers
reach consensus is not publicly documented.

\Cref{fig:dao-governance-bab} illustrates this scenario using the
visual grammar of \cref{sec:visual-grammar}.  Three actors within the
Blockchain at Berkeley community boundary each control an
\gls*{eoa}\@.  The community controls a \num{2}-of-\num{11}
multi-signature wallet that holds the COMP governance token and casts
a no~vote in Compound \gls*{dao}\@.  The diagram shows how a single
external community, operating through a multi-signature wallet, cast
the deciding vote on Proposal~\num{100}.

\begin{figure}[htbp]
  \centerline{\myincludegraphics[width=0.7\linewidth]{tikz/dao-governance/dao-governance-bab}}
  \caption{Blockchain at Berkeley's participation in Compound
    governance.  Three actors within the Blockchain at Berkeley
    community control a \num{2}-of-\num{11} multi-signature wallet
    that holds COMP tokens and casts a no~vote in Compound
  \gls*{dao}\@.}\label{fig:dao-governance-bab}
\end{figure}

\subsection{Index Coop}\label{sec:case-study-index-coop}

Index Coop simplifies access to complex DeFi strategies through ERC-20
tokens~\citep{index-coop-xx}, offering products such as tokenised
index funds (\unit{\dpi}).  A key distinction between traditional
financial stewardship and Index Coop's metagovernance model lies in
who makes the governance decisions.  In traditional finance,
decision-makers are portfolio managers, boards, and executives; in
Index Coop, governance rights are embedded in the \unit{\index} token,
which is separate from the underlying assets such as \unit{\dpi}.

We identified Index Coop's active participation in the governance of
four \glsstar{}{dao}{s} by counting the number of Index Coop proposals
that referenced each: Aave (\num{669} proposals), Compound
(\num{179}), Uniswap (\num{114}), and Balancer (\num{33}).  Index Coop
also voted on proposals in those \glsstar{}{dao}{s} directly:
\num{120} times on Aave, \num{32} times on Compound, twice on Uniswap,
and four times on Balancer.  Several factors account for the
discrepancy between Index Coop's internal proposals and its direct
votes.  First, Index Coop mandates a quorum for governance actions, so
a referenced proposal that fails to reach quorum produces no
corresponding vote.  Second, the referenced proposals may relate to
different stages of governance: Uniswap's on-chain proposals, for
example, follow \textquote{temperature check} Snapshot votes.  Third,
our heuristic may include false positives, where mentions of the four
\glsstar{}{dao}{s} were not direct metagovernance decisions.

Index Coop's involvement across four \glsstar{}{dao}{s} makes it a
clear instance of nexus metagovernance.

\section{Conclusions}\label{sec:conclusions}

\glsstar{}{dao}{s} promote transparency and auditability as
significant advantages over traditional governance frameworks.
However, metagovernance fragments the governance process, detaches
votes from their context, and introduces security risks.  In this
paper, we have presented a method for identifying external community
governance interactions, characterising the mechanics of
inter-\gls*{dao} governance through multi-signature smart contracts,
the forms in which metagovernance communities manifest, and the
influence external communities exert on the \glsstar{}{dao}{s} they
govern.

The detection method has two parts:
\begin{enumerate}
  \item \emph{On-chain governance contract identification}: starting
    from a voting power contract, we sort all contract accounts that
    interact with it by frequency and cross-reference their event
    signatures against a database of voting and proposal signatures.
  \item \emph{Contract account voter extraction}: from the identified
    governance contract, we collect the voting event log and extract
    all voter addresses.  We retain only those addresses that contain
    executable code, i.e., contract accounts, since these are the
    candidates for external community participation.
\end{enumerate}

For off-chain metagovernance, we matched a \glsstar{}{dao}{'s}
on-chain voting power contracts to Snapshot Spaces using a database
constructed from Snapshot's GraphQL \glstext*{api}\@.  When multiple
Spaces referenced the same voting power contract, we selected the most
followed Space, excluding any with fewer than \num{50} followers.  We
then parsed the voting data from the Snapshot \glstext*{api} and
retained only voters that were contract accounts.

We labelled contract accounts using Etherscan's public name tags,
supplemented by names and comments in the contract source code.  This
labelling allowed us to identify the entities behind metagovernance
interactions and to expand the dataset by following labelled edges.
The result is a metagovernance network that visualises the
metagovernance relationships between \glsstar{}{dao}{s}.

Across \num{61} established \glsstar{}{dao}{s}, we identified \num{72}
external community relationships.  These relationships are a detected
subset rather than a complete census of metagovernance on Ethereum:
our heuristics did not capture every \gls*{dao} framework, as shown by
the detection misses reported in \cref{sec:results}, and further
relationships may exist both within and beyond the \glsstar{}{dao}{s}
we examined.  We selected three of these relationships for closer
examination, and they exhibit different forms: from a university group
operating a multi-signature wallet (Blockchain at Berkeley), through a
single protocol acquiring delegated voting power to list its own token
(Liquity), to an entire \gls*{dao} with its own token holders that
replicates the underlying \glsstar{}{dao}{'s} proposals on its own
governance platform (Index Coop).

The impact of these external communities also varies across the three
cases.  Blockchain at Berkeley cast the deciding vote on a Compound
proposal that would otherwise have passed, demonstrating that a single
external community's vote can determine the outcome.  Liquity proposed
and voted to list its own token on the underlying \glsstar{}{dao}{'s}
lending platform, using delegated voting power to advance its own
protocol's interests.  Index Coop relocated the governance process of
the underlying \glsstar{}{dao}{s} by replicating their proposals on
its own governance platform, where votes are cast using the
\unit{\index} token rather than the underlying \glsstar{}{dao}{'s}
token.

Although these three cases are a small subset of the relationships in
our dataset, together they suggest three concerns.  First, external
communities can be decisive in the governance outcomes of the
\glsstar{}{dao}{s} they govern, indicating that they hold real power
within those \glsstar{}{dao}{s}.  Second, they may use that voting
power to directly benefit their own community, introducing potential
conflicts of interest with the underlying community.  Third, external
community governance can relocate the locus of voting from the
underlying \glsstar{}{dao}{'s} governance contract to another smart
contract operating under a different governance framework, reducing
transparency, which is fundamental to \gls*{dao} governance.

Despite successes in identifying \gls*{dao}-to-\gls*{dao} voting, the
on-chain metagovernance algorithm has flaws.  We suggest three areas
for improvement in future iterations:
\begin{enumerate}
  \item Improve the identification of voting power relationships:
    governance tokens and their associated \glsstar{}{dao}{s} may have
    separate voting power contracts not accounted for in our current
    method.  An additional algorithm could identify these abstracted
    relationships between tokens and voting power contracts.
  \item Reduce false positives: our current algorithm for on-chain
    governance detection can generate false positives.  Further
    iterations should focus on identifying and filtering out wrapper
    contracts to improve accuracy.
  \item Expand the index for contract labelling: the current reliance
    on name tags and manual searches for contract context is limiting.
    Future work should integrate additional labelling sources, such as
    contract source code repositories, project documentation, and
    cross-chain identity registries, to improve coverage and reduce
    the manual effort required to interpret unlabelled accounts.
\end{enumerate}

Active external \gls*{dao} communities exist throughout Ethereum's
prominent \glsstar{}{dao}{s}.  They are not uniform: they range from a
handful of signers controlling a multi-signature contract to entire
protocols with hundreds of token holders.  The interconnected
\gls*{dao} governance ecosystem currently operates without major
disruption, but it creates conditions for inter-\gls*{dao} friction
that the underlying \glsstar{}{dao}{s} neither authorised nor control.

\glsstar{}{dao}{s} govern protocol upgrades, treasury management, and
other consequential matters.  Token holders within a single \gls*{dao}
typically share aligned incentives for its success; metagovernance
introduces external parties whose incentives may not be aligned with
the underlying \gls*{dao}\@.  Because blockchain governance lacks the
regulatory and social checks present in corporate governance, the
outcomes of metagovernance relationships are difficult to anticipate.
Tools that identify and contextualise metagovernance, of the kind we
have begun to develop in this paper, are a necessary step towards
understanding the governance dynamics that emerge across \gls*{dao}
boundaries.

\bibliographystyle{my-abbrvnat}
\bibliography{metagovernance}

\end{document}